\documentstyle[aps,prb,preprint]{revtex} 

\begin{document}  \preprint{\today} \draft
\title{Hole Photoproduction in Insulating
Copper Oxide}

\author{O. P. Sushkov\cite{UNSW,Budker}, G. Sawatzky,
R. Eder, and H. Eskes}

\address{Materials Science Center, University of Groningen\\
Nijenborgh 4, 9747 AG Groningen\\ The Netherlands}
\maketitle

\begin{abstract}
To explain the experimental spectra for angle resolved photoemission
we consider a modified $t-J$ model. The modified model includes
next nearest ($t^{\prime}$) and next next nearest ($t^{\prime \prime}$)
hopping as well as Hubbard model corrections to the spectral weights.A 
Dyson equation which relates the single hole Green's functions for a given
pseudospin and given spin is derived and is compared to experimental results.
\end{abstract}

\narrowtext 
\section{Introduction}
Recent angle resolved photoemission (ARPES) measurements by Wells 
{\it et al} \cite{Wells} and by Pothuizen {\it et al}\cite{Pothuizen}
for insulating Copper Oxide Sr$_2$CuO$_2$Cl$_2$ provide a  unique possibility 
to experimentally determine the single hole properties. 
In the  frameworks of a  $t-t^{\prime}-J$ model the problem has been analyzed 
by Nazarenko {\it et al} \cite{Naz} using a  cluster method and
Bala, Oles, and Zaanen\cite{Bala} using a  self-consistent Born approximation
(SCBA). It was demonstrated in these works that including $t^{\prime}$ hopping
improves agreement with experiment, but a  complete description of the 
ARPES spectra especially the k dpendent intensity was not achieved.

In the  present work we consider a  $t-t^{\prime}-t^{\prime \prime}-J$
model. A  Dyson equation which relates the one  electron Green's function measured 
in experiment to the  hole Green's function found in the self-consistent Born 
approximation is derived. We also introduce into this equation the 
corrections which originate from finite U as in the Hubbard model. 
With parameters taken from LDA band calculations by Andersen {\it et al} 
\cite{And}
we got a reasonably good  description of the experimental ARPES spectra . 
The results are sensitive to $t^{\prime \prime}$. The parameter 
$t^{\prime}$ is less important.
The importance of $t^{\prime \prime}$ was also pointed out
by Belinicher and Chernyshev \cite{Cher}.
For the explanation of width of ARPES spectra we need to
introduce additional broadening and possible origins of this
are discussed .

The Hamiltonian of $t-t^{\prime}-t^{\prime \prime}-J$ model is
of the form
\begin{equation}
\label{H}
H=-t\sum_{\langle ij \rangle \sigma} c_{i\sigma}^{\dag}c_{j\sigma}
-t^{\prime}\sum_{\langle ij_1 \rangle \sigma}
c_{i\sigma}^{\dag}c_{j_1\sigma}
-t^{\prime \prime}\sum_{\langle ij_2 \rangle \sigma} 
c_{i\sigma}^{\dag}c_{j_2\sigma}
+J \sum_{\langle ij \rangle \sigma} {\bf S}_i{\bf S}_j.
\end{equation}
$c_{i \sigma}^{\dag}$ is the  creation operator of an electron with
spin $\sigma$ $(\sigma =\uparrow, \downarrow)$ at site $i$
of the two-dimensional square lattice,the $\langle ij \rangle$ represents
nearest neighbor sites, $\langle ij_1 \rangle$ - next nearest neighbor
(diagonal), and $\langle ij_2 \rangle$ represents next next
nearest sites. The spin operator is ${\bf S}_i={1\over 2}
c_{i \alpha}^{\dag} {\bf \sigma}_{\alpha \beta} c_{i \beta}$.
The size of the exchange measured in two magnon Raman scattering
\cite{Tok,Grev} is $J=125meV$. The most recent calculation of the
hopping matrix elements has been done by Andersen {\it et al}
\cite{And}. They consider a  two-plane situation and the effective
matrix elements are slightly different for symmetric and
 antisymmetric combinations of orbitals between planes. After averaging
over these combinations we get: $t=386meV$, $t^{\prime}=-105meV$,
$t^{\prime \prime}=86meV$. Below we set $J=1$, in these units
\begin{equation}
\label{ts}
t=3.1, \ \ \ t^{\prime}=-0.8, \ \ \ t^{\prime \prime}=0.7
\end{equation}

\section{Hole Green's function with fixed pseudospin.
Self-consistent Born approximation (SCBA)}
It is well known that at half filling (one electron per site)
the model under consideration is equivalent to a  Heisenberg
model. It represents a  Mott insulator with long range antiferromagnetic
order. We denote the corresponding ground state wave function by $|0\rangle$.
We are interested in the situation when one electron is
removed from this state, so a single hole is produced.
The dynamics of a single hole in the antiferromagnetic background
can be described by SCBA
\cite{Schmitt,Kane}. This approximation works very well due to the
absence of a single loop correction to the hole-spin-wave vertex
\cite{Mart,Liu,Susf}. Let us recall the idea of this approximation.
The bare hole operator $d_i$ is defined so that $d_{i}^{\dag}
\propto c_{i \uparrow}$ on the $\uparrow$ sublattice and 
$\propto c_{i \downarrow}$ 
on the $\downarrow$ sublattice. In the momentum representation
\begin{equation}
\label{d}
d_{{\bf k}\downarrow}^{\dag} =\sqrt{2\over {N (1/2+m)}}\sum_{i \in \uparrow}
c_{i \uparrow}e^{i{\bf k}{\bf r}_i}, \ \ \ \
d_{{\bf k}\uparrow}^{\dag} =\sqrt{2\over {N (1/2+m)}}\sum_{j \in \downarrow}
c_{j \downarrow}e^{i{\bf k}{\bf r}_j}.
\end{equation}
$N$ is number of sites, $m= |\langle 0|S_{iz}|0\rangle|\approx 0.3$ is
the average sublattice magnetization . The quasi-momentum
${\bf k}$ is limited to be  inside the magnetic
Brillouin zone: $\gamma_{\bf k}= {1\over 2}(\cos k_x + \cos k_y) \ge 0$.
In this notations it looks like $d_{{\bf k} \sigma}$ has spin
$\sigma =\pm 1/2$, but actually rotation invariance is violated
and $\sigma$ is a  pseudospin which denotes the sublattice. Nevertheless
the pseudospin gives the correct value of the spin $z$-projection:
$S_z=\sigma =\pm 1/2$. The coefficients in (\ref{d}) provide the correct
normalization:
\begin{equation}
\label{norm}
\langle 0|d_{{\bf k}\downarrow} d_{{\bf k}\downarrow}^{\dag}|0\rangle
={2\over {N (1/2+m)}}\sum_{i \in \uparrow}
\langle 0|c_{i \uparrow}^{\dag}c_{i \uparrow}|0\rangle
={1\over {1/2+m}}\langle 0|{1\over 2}+S_{iz}|0\rangle=1.
\end{equation}
The hole Green's function is defined as
\begin{equation}
\label{Gd}
G_d(\epsilon,{\bf k})=-i\int \langle 0|T d_{{\bf k}\sigma}(\tau)
d_{{\bf k}\sigma}^{\dag}(0)|0 \rangle e^{i \epsilon \tau} d\tau
\end{equation}
The $t^{\prime}$, $t^{\prime \prime}$  terms
in the Hamiltonian (\ref{H}) correspond to the 
hole hopping inside one sublattice. This gives the bare hole dispersion
\begin{equation}
\label{e0}
\epsilon_{0{\bf k}}=4t^{\prime}\cos k_x \cos k_y+
2t^{\prime \prime}(\cos 2k_x + \cos 2k_y) \to
\beta_{01}\gamma^2_{\bf k}+\beta_{02}(\gamma^-_{\bf k})^2,
\end{equation}
where $\gamma^-_{\bf k}={1\over 2}(\cos k_x - \cos k_y)$,
$\beta_{01}=4(2t^{\prime \prime}+t^{\prime})$, and
$\beta_{02}=4(2t^{\prime \prime}-t^{\prime})$.
In equation (\ref{e0}) we took into account that the sign of a hole
dispersion is opposite to that for an electron (maximum of electron
band correspond to minimum of hole band), and omitted some constant. 
The bare hole Green's function is
\begin{equation}
\label{Gd0}
G_{0d}(\epsilon, {\bf k})={1\over{\epsilon - \epsilon_{0{\bf k}}+i0}}.
\end{equation}

For spin excitations the usual linear spin-wave theory is used
(see, e.g. review paper\cite{Manousakis}). It is convenient
to have two types of spin-waves, $\alpha_{\bf q}^{\dag}$ with 
$S_z=-1$, and $\beta_{\bf q}^{\dag}$ with $S_z=+1$, and ${\bf q}$
restricted to be  inside the magnetic Brillouin zone. 
\begin{eqnarray}
\label{swt}
\sqrt{2\over N}\sum_{i \in \uparrow}S_i^+e^{-i{\bf q r_i}}&\approx&
u_{\bf q}\alpha_{\bf q}+v_{\bf q}\beta_{\bf -q}^{\dag},\\
\sqrt{2\over N}\sum_{j \in \downarrow}S_j^-e^{i{\bf q r_j}}&\approx&
v_{\bf q}\alpha_{\bf q}^{\dag}+u_{\bf q}\beta_{\bf -q}.\nonumber
\end{eqnarray}
The spin-wave dispersion
and parameters of Bogoliubov transformation diagonalizing the
spin-wave Hamiltonian are:
\begin{eqnarray}
\label{sw}
&&\omega_{\bf q}=2\sqrt{1-\gamma_{\bf q}^2},\nonumber \\
&&u_{\bf q}=\sqrt{{1\over{\omega_{\bf q}}}+{1\over 2}},\\
&&v_{\bf q}=-sign(\gamma_{\bf q})
\sqrt{{1\over{\omega_{\bf q}}}-{1\over 2}}.\nonumber
\end{eqnarray}

Hopping to nearest a  neighbor in the Hamiltonian (\ref{H}) gives an
interaction of the hole with spin-waves. 
\begin{equation}
\label{hsw}
H_{h,sw}=\sum_{\bf k,q}g_{\bf k,q} \left(
d_{{\bf k+q}\downarrow}^{\dag}d_{{\bf k}\uparrow}\alpha_{\bf q}+
d_{{\bf k+q}\uparrow}^{\dag}d_{{\bf k}\downarrow}\beta_{\bf q}+
H.c.\right),
\end{equation}
with vertex $g_{\bf k,q}$ given by
\begin{eqnarray}
\label{g}
g_{\bf k,q} &\equiv& \langle 0|\alpha_{\bf q}d_{{\bf k}\uparrow}|H_t|
d_{{\bf k+q}\downarrow}^{\dag}|0\rangle=\nonumber\\
&=&{2\over {N (1/2+m)}}
\langle 0|\alpha_{\bf q}\sum_{\langle i \in \uparrow,
j \in \downarrow \rangle}
e^{-i{\bf k}{\bf r}_j}c_{j \downarrow}^{\dag}\left(-t c^{\dag}_{i\sigma}
c_{j\sigma}\right)c_{i \uparrow}e^{i({\bf k+q}){\bf r}_i}|0\rangle=\nonumber\\
&=&{2\over {N (1/2+m)}}t\sum_{\langle i \in \uparrow,
j \in \downarrow \rangle}e^{-i{\bf k}{\bf r}_j+i({\bf k+q}){\bf r}_i}
\left(
\langle 0|\alpha_{\bf q}c_{j \downarrow}^{\dag}c_{j \uparrow}
c_{i \uparrow}^{\dag}c_{i \uparrow}|0\rangle+
\langle 0|\alpha_{\bf q}c_{j \downarrow}^{\dag}c_{j \downarrow}
c_{i \downarrow}^{\dag}c_{i \uparrow}|0\rangle\right)\approx\nonumber\\
&\approx&{2\over N}t\sum_{\langle i \in \uparrow,
j \in \downarrow \rangle}e^{-i{\bf k}{\bf r}_j+i({\bf k+q}){\bf r}_i}
\langle 0|\alpha_{\bf q}\left(S_j^- +S_i^-\right)|0\rangle=
4t\sqrt{2\over N}(\gamma_{\bf k}u_{\bf q}+
\gamma_{\bf k+q}v_{\bf q}).
\end{eqnarray}
In this calculation we have used the usual mean field ground state
factorization approximation: 
$\langle 0|\alpha_{\bf q}c_{j \downarrow}^{\dag}c_{j \uparrow}
c_{i \uparrow}^{\dag}c_{i \uparrow}|0\rangle \approx
\langle 0|\alpha_{\bf q}c_{j \downarrow}^{\dag}c_{j \uparrow}|0\rangle
\langle 0|c_{i \uparrow}^{\dag}c_{i \uparrow}|0\rangle=
\langle 0|\alpha_{\bf q}S_j^-|0\rangle (1/2+m)$.
The vertex $g_{\bf k,q}$ is independent of $t^{\prime}$,
$t^{\prime \prime}$ because these parameters correspond to hopping inside
one sublattice. The form of the vertex $g_{\bf k,q}$ is well known.
The actual purpose of calculation (\ref{g}) is to demonstrate that
$g_{\bf k,q}$ is valid in a more general situation than it is
usually believed. It is independent of the particular value of the
sublattice magnetization $m$. Therefore, for example, $g_{\bf k,q}$
remains the same in the presence of strong additional frustrations.

One can easily prove that the spin structure of the 
interaction (\ref{hsw}) forbids single loop corrections  to the
hole-spin-wave vertex and, as usually, the two loop correction is small
numerically\cite{Mart,Liu,Susf}. So, due to the spin
structure we have an analog of the well known Migdal theorem for
electron-phonon interactions. This justifies SCBA according to which the
hole Green's function satisfies a  simple Dyson equation
\begin{equation}
\label{Dy}
G_d(\epsilon, {\bf k})=\left(\epsilon - \epsilon_{0{\bf k}}
-\sum_{\bf q}g^2_{\bf k-q,q} G_d(\epsilon -\omega_{\bf q},{\bf k-q})
+ i 0 \right)^{-1}.
\end{equation}
The anomalous  Green's function $-i \langle 0|T d_{{\bf k}\uparrow}(t)
d_{{\bf k}\downarrow}^{\dag}(0)|0 \rangle$ vanishes because the
$z$-projection of spin is conserved.
Due to the definition of the operators (\ref{d}) the Green's function
(\ref{Gd}) is invariant under translation  with the inverse vector
of the magnetic sublattice ${\bf Q}=(\pm \pi,\pm \pi)$
\begin{equation}
\label{Bl}
G_d(\epsilon,{\bf k+Q})=G_d(\epsilon,{\bf k}).
\end{equation}

The numerical solution of equation (\ref{Dy}) is straightforward.
As usual, to avoid poles we replace $i0 \to i \Gamma/2 =i \ 0.1$. 
The energy scale consists of 300 points with variable density
(concentrated near sharp structures of $G_d$).
The number of points in the magnetic Brillouin zone is $10^4$
which is equivalent to the lattice $140 \times 140$.
The plots of $-{1\over{\pi}} \ Im \ G_d(\epsilon, {\bf k})$
as a functions of $\epsilon$ for ${\bf k}=(\pi/2, \pi/2)$,
${\bf k}=(\pi/2, 0)$, ${\bf k}=(\pi,0)$, and ${\bf k}=(0,0)$
are presented in Fig.1. We recall that we use the set of parameters 
(\ref{ts}) based on Ref.\cite{And}. 
The position of the lowest peak gives the quasiparticle energy.
Results of the calculation can be fitted by the formula
\begin{eqnarray}
\label{e2}
\epsilon_{\bf k}& = & const+
\beta_1\gamma^2_{\bf k}+\beta_2(\gamma^-_{\bf k})^2
+\beta_2^{\prime}(\gamma^-_{\bf k})^4\\
&&\beta_1 \approx 3.0, \ \ \ \beta_2 
\approx 3.8, \ \ \ \beta_2^{\prime} \approx -1.5. \nonumber
\end{eqnarray}
The dispersion has  minima at (${\bf k_0}=(\pm \pi/2, \pm \pi/2)$.
The hole pockets are slightly stretched along the direction
to the zone center, and it is very different from a  pure $t-J$ 
model \cite{Mart,Liu}. The quasiparticle residue $Z_{\bf k}^d$ can be
found as the area under the peak. At the dispersion minimum it equals 
$Z_{\bf k_0}^d=0.38$. 
So it is larger than in a  pure $t-J$ model\cite{Mart,Liu},
but away from the dispersion minimum it drops down very rapidly.
Plots of the residue $Z_{\bf k}^d$ as a function of ${\bf k}$
for ${\bf k} \in \left[(\pi/2,\pi/2) \ - \ (0,0)\right]$
and ${\bf k} \in \left[(\pi/2,\pi/2) \ - \ (0,\pi)\right]$
are given at Fig. 2. Throughout the Brillouin zone the residue is
fitted by
\begin{equation}
\label{Z}
Z_{\bf k}^d=Z_{\bf k_0}^d\left(1-0.9\gamma^2_{\bf k}-
1.52(\gamma^-_{\bf k})^2
+0.05\gamma^4_{\bf k}+0.69(\gamma^-_{\bf k})^4
+0.5(\gamma_{\bf k}\gamma^-_{\bf k})^2\right).
\end{equation}

\section{Hole Green's function with fixed spin.
Dyson equation relating two Green's functions}
The operators $d_{{\bf k}\uparrow}$, $d_{{\bf k}\downarrow}$ 
discussed in the previous section are
defined at different sublattices. However, when a photon kicks out
an electron from the system it does not separate the sublattices.
Therefore for this process we have to define the operator as a simple
Fourier transform
\begin{equation}
\label{c}
c_{{\bf k}\sigma} =\sqrt{2\over N}\sum_i
c_{i \sigma}e^{i{\bf k}{\bf r}_i}.
\end{equation}
The normalization is chosen in such a way that 
$\langle 0|c^{\dag}_{{\bf k}\uparrow} c_{{\bf k}\uparrow}|0\rangle=
{2 \over N} \langle 0|\sum_i c^{\dag}_{i\uparrow}c_{i\uparrow} 
|0\rangle=1$. We can consider $c_{{\bf k}\sigma}$ as an external perturbation,
and the corresponding Green's function is
\begin{equation}
\label{Gc}
G_c(\epsilon,{\bf k})=-i\int \langle 0|T c_{{\bf k}\sigma}^{\dag}
(\tau)c_{{\bf k}\sigma}(0)|0 \rangle e^{i \epsilon \tau} d\tau.
\end{equation}
This is the Green's function measured in ARPES. 
Let us now find the relation between $G_c(\epsilon,{\bf k})$
and $G_d(\epsilon,{\bf k})$.

The operator $c_{{\bf k}\sigma}$ acting on the vacuum (ground state
of the Heisenberg model) can produce a single hole state. We denote the
corresponding amplitude by $a_{\bf k}$ and  show it in  Fig. 3a
as a cross.The thick line corresponds to the Green's function $G_c$ (\ref{Gc})
and the thin line corresponds to the $G_d$ (\ref{Gd}). The amplitude
$a_{\bf k}$ equals
\begin{equation}
\label{ak}
a_{\bf k}=\langle 0| d_{{\bf k}\uparrow} c_{{\bf k}\downarrow}|0\rangle
=\langle 0|\left(
\sqrt{2\over {N (1/2+m)}}\sum_{j \in \downarrow}
c_{j \downarrow}^{\dag}e^{-i{\bf k}{\bf r}_j}\right)
\left(\sqrt{2\over N}\sum_{i}
c_{i \downarrow}e^{i{\bf k}{\bf r}_i}\right)|0\rangle =\sqrt{1/2+m}.
\end{equation}
The operator $c_{{\bf k}\sigma}$ acting on the vacuum  can also produce a 
hole + spin-wave state. This amplitude is shown in Fig. 3b
as a circled cross with the dashed line being a spin-wave. We denote this
amplitude by $b_{\bf k,q}$  
\begin{eqnarray}
\label{bk}
b_{\bf k,q}&=&\langle 0|\beta_{\bf q} 
d_{{\bf k-q}\downarrow} c_{{\bf k}\downarrow}|0\rangle
=\langle 0|\beta_{\bf q} \left(\sqrt{2\over {N (1/2+m)}}\sum_{i \in \uparrow}
c_{i \uparrow}^{\dag}e^{-i({\bf k-q}){\bf r}_i}\right)
\left(\sqrt{2\over N}\sum_{j}
c_{j \downarrow}e^{i{\bf k}{\bf r}_j}\right)|0\rangle\approx\nonumber\\
&\approx&{2\over{N}}\langle 0|\beta_{\bf q}\left( \sum_{i \in \uparrow}
S_i^+ e^{i{\bf qr_i}}\right)|0\rangle = \sqrt{2\over{N}} v_{\bf q}.
\end{eqnarray}
We stress that (\ref{bk}) is a bare vertex. It corresponds to the
instantanious creation of a hole + spin wave, but not the creation of a hole
with a  subsequent decay into hole + spin-wave. 
To elucidate this point look at Fig.4.The  upper part of this figure
describes the wave function of the initial Neel state: a - component without 
spin quantum fluctuations, b - component with spin quantum fluctuation. 
The lower part of Fig.4 arises
imediately from the upper one after kicking out  an electron with spin up. 
Part a does not contain a  spin flip, and it corresponds to the amplitude 
$a_{\bf k}$ (\ref{ak}). Part b does contain a spin flip, and it corresponds to 
the amplitude $b_{\bf k,q}$ (\ref{bk}). Note that $b_{\bf k,q}
\to \infty$ at ${\bf q} \to 0$. The reason is that the operator
(\ref{c}) does not correspond to any quasiparticle of the system
with long-range antiferromagnetic order, and therefore the usual Goldstone 
theorem is not applicable.

Let us denote by a dot (Fig. 3c) the usual hole-spin-wave vertex 
$g_{\bf k,q}$ given by eq. (\ref{g}). In the leading in $t$
approximation the amplitude  of  single hole creation by the external 
perturbation (\ref{c}) is given by diagrams presented at Fig. 5,
with the thin solid line  in this case the bare hole Green's function
(\ref{Gd0}). If we set $t^{\prime}=t^{\prime \prime}=0$ and 
$\epsilon =\epsilon_{0{\bf k}}=0$
calculation of this amplitude can be easily done analytically
\begin{eqnarray}
\label{M1}
M_1(\epsilon =0,{\bf k})&=&
a_{\bf k}+\sum_{\bf q}{{b_{\bf k,q} g_{\bf k-q,q}}\over
{\epsilon_{0{\bf k}}-\epsilon_{0{\bf k-q}}-\omega_{\bf q}}}=
\sqrt{0.8}-{{8t}\over{N}}\sum_{\bf q}
{{v_{\bf q} (\gamma_{\bf k-q} u_{\bf q}
+\gamma_{\bf k}v_{\bf q})}\over{\omega_{\bf q}}}=\nonumber\\
&=&\sqrt{0.8}+{{4t}\over{N}}\gamma_{\bf k}\sum_{\bf q}
\left({1\over{\omega_{\bf q}}}-
{1\over 2}\right)=\sqrt{0.8}\left(1+0.45\cdot t \cdot \gamma_{\bf k}\right).
\end{eqnarray}
$M_1^2$ is the quasiparticle residue of the Green's function (\ref{Gc}).
The eq. (\ref{M1}) agrees with the result obtained
using a  string representation\cite{Eder}. We stress that even at $t=0$
the residue is 0.8 due to the spin quantum fluctuation in the ground state 
of the Heisenberg model\cite{Mal,Poil93}

Now we can find the relation between Green's functions $G_c$ (\ref{Gc}) 
and $G_d$ (\ref{Gd}). In the leading in $t$ approximation it is
given by diagrams presented at Fig. 6 with the thin solid line being in this
case the bare hole Green's function $G_{0d}$ (\ref{Gd0}).
Now let us dress these diagrams by higher orders in hopping $t$.
As we already discussed there is no single loop correction to the 
``dot''.  We neglect double loop correction to the ``dot''  as it has 
been done in  SCBA. Therefore the only possibility is an introduction of
a self energy corrections. An example of the correction to
diagram Fig. 6c is shown at Fig. 7.
To take into account all these corrections we need just to replace
at Fig. 6 all bare hole Green's functions (\ref{Gd0}) by dressed
hole Green's function given by eq. (\ref{Dy}).
So, the Fig. 6 actually represents a  Dyson equation relating $G_c$ (\ref{Gc}) 
and $G_d$ (\ref{Gd}). In analytical form it is
\begin{eqnarray}
\label{Dy2}
G_c(\epsilon,{\bf k})&=&a_{\bf k}^2 G_d(\epsilon,{\bf k})
+\sum_{\bf q}b_{\bf k,q}^2 G_d(\epsilon-\omega_{\bf q},{\bf k-q})+
\nonumber \\
&+&2 a_{\bf k} G_d(\epsilon,{\bf k})\left[
\sum_{\bf q}b_{\bf k,q} g_{\bf k-q,q} G_d(\epsilon-\omega_{\bf q},{\bf k-q})
\right]+\\
&+&G_d(\epsilon,{\bf k}) \left[
\sum_{\bf q}b_{\bf k,q} g_{\bf k-q,q} G_d(\epsilon-\omega_{\bf q},{\bf k-q})
\right]^2.\nonumber
\end{eqnarray}
So as soon as we have found $G_d$ using SCBA
(\ref{Dy}) we can calculate the Green's function $G_c$ defined by eq. (\ref{Gc}).
The imaginary part of $G_c(\epsilon,{\bf k})$ gives directly the spectra
measured in ARPES experiments.

We now  discuss sum rules. All singularities of Green's functions
are in the lower half plane of complex $\epsilon$. Therefore if
we integrate eq.(\ref{Dy}) over $\epsilon$ from $-\infty$ to
$+\infty$, this integral can be replaced by the integral over an
infinite semi-circle in the upper $\epsilon$ half plane.For infinite
$\epsilon$, $G_d = G_{0d}$, and we get the well known sum rule
\begin{equation}
\label{sum}
-{1\over{\pi}} Im \int_{-\infty}^{\infty}G_d(\epsilon,{\bf k})
d \epsilon =1,
\end{equation}
which agrees with with eq.(\ref{norm}).
If we integrate now eq.(\ref{Dy2}) in the same limits, the terms
which contain more than one Green's function give zero contribution,
because the integral can be transfered into the upper complex
$\epsilon$ half plane. And we find
\begin{equation}
\label{sum1}
-{1\over{\pi}} Im \int_{-\infty}^{\infty}G_c(\epsilon,{\bf k})
d \epsilon =\left(
-{1\over{\pi}} Im \int G_d(\epsilon,{\bf k})
d \epsilon \right) \left(a_{\bf k}^2+ \sum_{\bf q}b_{\bf k,q}^2\right)=
0.8+{2\over{N}}\sum_{\bf q} v_{\bf q}^2 = 1.
\end{equation}
Thus the equation (\ref{Dy2}) reproduces the correct normalization:
$\langle 0|c^{\dag}_{{\bf k}\uparrow} c_{{\bf k}\uparrow}|0\rangle=1$. 

The vertex $b_{\bf k,q}$ (\ref{bk}) is invariant under translation with 
the inverse vector of magnetic sublattice ${\bf Q}=(\pm \pi,\pm \pi)$:
$b_{\bf k+Q,q}=b_{\bf k,q}$. At the same time the vertex $g_{\bf k,q}$
(\ref{g}) changes  sign with this translation:
$g_{\bf k+Q,q}=-g_{\bf k,q}$. Therefore the diagrams Fig. 6c,d
change sign at ${\bf k} \to {\bf k+Q}$ and
\begin{equation}
\label{ne}
G_c(\epsilon,{\bf k+Q}) \ne G_c(\epsilon,{\bf k}).
\end{equation}
Due to the same properties of vertices $b_{\bf k,q}$ and $g_{\bf k,q}$
the diagrams Fig. 6c,d,e (square brackets in eq. (\ref{Dy2})) vanish 
at the face of magnetic Brillouin zone ($\gamma_{\bf k}=0$).
The diagram presented at Fig. 6b (term with $b_{\bf k,q}^2$ in eq.
(\ref{Dy2})) is small numerically. Therefore at the face of magnetic
Brillouin zone $G_c(\epsilon,{\bf k}) \approx G_d(\epsilon,{\bf k})$.
However away from the face they differ significantly.
The plots of $-{1\over{\pi}} \ Im \ G_c(\epsilon, {\bf k})$
as a functions of $\epsilon$ for ${\bf k}=(\pi/2, \pi/2)$,
${\bf k}=(\pi/2,0)$, ${\bf k}=(\pi,0)$, and ${\bf k}=(0,0)$
are presented at Fig.8.
A plot of the quasiparticle residue $Z_{\bf k}^c$ as a function
of ${\bf k}$ along (1,1) direction is given at Fig. 9.
The quasiparticle residue outside the magnetic zone is smaller than
that inside. For comparison we also present a  plot of $Z_{\bf k}^d$.

\section{Hubbard model correction}
The picture considered above corresponded to a modified $t-J$ model. It
means that double electron occupancy was forbidden. 
Now we want to take into account the fact that the $t-t^{\prime}-
t^{\prime \prime}-J$ model originates from the Hubbard model.
We assume that it is a simple one band Hubbard model with on site
repulsion $U$.
First of all this gives some corrections to the ``bare'' hole dispersion
(\ref{e0}), see, e. g. Ref.\cite{Bala}. However we assume that 
renormalization is done and these corrections are already included 
in the values of effective hopping amplitudes $t^{\prime}$, 
$t^{\prime \prime}$ given in (\ref{ts}).
There are also some corrections to the hole-spin-wave vertex\cite{Bala},
but they are small at $t^{\prime},t^{\prime \prime} \ll U$.
The really important effect is the renormalization of the vertex 
$a_{\bf k}$ (\ref{ak}). In $t-J$ model this vertex is given by
the process shown at Fig. 10a: an electron is removed from
corresponding sublattice. In Hubbard model there is an additional
possibility shown at Fig. 10b: first the electron hops to occupied
nearest site, and then it is removed from this site. Simple
calculation shows that this gives
\begin{eqnarray}
\label{ak1}
a_{\bf k}&\to& a_{\bf k}\times \left(1+{{4t}\over{U}}\gamma_{\bf k}\right))=
\sqrt{1/2+m}\left(1+{{J}\over{t}}\gamma_{\bf k}\right),\\
b_{\bf k,q}&\to& b_{\bf k,q}\times \left(1+{{4t}\over{U}}\gamma_{\bf k}\right)=
\sqrt{2\over{N}} v_{\bf q}\left(1+{{J}\over{t}}\gamma_{\bf k}\right).
\nonumber
\end{eqnarray}
We took into account that $J=4t^2/U$. The magnitude of the $t/U$ correction
in (\ref{ak1}) is obvious, however one should be careful with the sign.
To find it one needs to commute fermionic operators in order corresponding 
to Fig. 10b. The Dyson equation (\ref{Dy2}) remains valid.
So we can easily find the Green's function $G_c^H$,
where index $H$ indicates that the Hubbard model correction is
taken into account.
The plots of $-{1\over{\pi}} \ Im \ G_c^H(\epsilon, {\bf k})$
as functions of $\epsilon$ for ${\bf k}=(\pi/2, \pi/2)$,
${\bf k}=(\pi/2,0)$, ${\bf k}=(\pi,0)$, and ${\bf k}=(0,0)$
are presented in Fig.11.
A plot of the quasiparticle residue $Z_{\bf k}^{cH}$ as a function
of ${\bf k}$ along (1,1) direction is given in Fig. 9.
We see that the ``Hubbard correction'' causes the decrease of the residue
outside of the magnetic Brillouin zone to be steeper.

The sum rule (\ref{sum1}) is changed. Now we have 
\begin{equation}
\label{sum2}
-{1\over{\pi}} Im \int_{-\infty}^{\infty}G_c^H(\epsilon,{\bf k})
d \epsilon \approx \left(1+{{J}\over{t}}\gamma_{\bf k}\right)^2
\approx 1+2{{J}\over{t}}\gamma_{\bf k}.
\end{equation}
Let us comment on the definition (\ref{c}) of the operator 
$c_{{\bf k}\sigma}$. Its normalization is adjusted for a system
with strong antiferromagnetic correlations and it is close
to that for $d_{{\bf k}\sigma}$ (see eq.(\ref{d})). However as a
result the definition (\ref{c}) differs from that usually accepted 
for a normal Fermi liquid by a factor $\sqrt{2}$. This is the reason
why the sum rule (\ref{sum2}) can be larger than unity. 
Generally the normalization can be chosen arbitrally. It is a question
of convenience only. However, let us  prove that the sum rule for the total 
number of electrons in the system is fulfilled.
According to definition (\ref{c})
\begin{equation}
\label{den}
c_{{\bf k}\sigma}^{\dag}c_{{\bf k}\sigma}=2 N_{{\bf k}\sigma},
\end{equation}
where
\begin{equation}
\label{numb}
N_{{\bf k}\sigma}=\left(\sqrt{1\over N}\sum_i
c_{i \sigma}^{\dag}e^{-i{\bf k}{\bf r}_i}\right)
\left(\sqrt{1\over N}\sum_j
c_{i \sigma}e^{i{\bf k}{\bf r}_j}\right)
\end{equation}
is the operator for the number of electrons.
Due to the definition (\ref{Gc}) of Green's function $G_c$ one has 
the standard relation
\begin{equation}
\label{sum3}
-{1\over{\pi}} Im \int_{-\infty}^{\infty}G_{c,\sigma}^H(\epsilon,{\bf k})
d\epsilon
=\langle 0|c_{{\bf k}\sigma}^{\dag}c_{{\bf k}\sigma}|0\rangle=
2 \langle 0|N_{{\bf k}\sigma}|0\rangle.
\end{equation}
Comparing with (\ref{sum2}) we find
\begin{equation}
\label{nn}
\langle 0|N_{{\bf k}\sigma}|0\rangle = {1\over 2}
(1+2{{J}\over{t}}\gamma_{\bf k}).
\end{equation}
The operator for the total number of electrons is equal to
\begin{equation}
\label{totn}
\hat{N}=\sum_{\sigma, {\bf k}\in full}N_{{\bf k}\sigma}.
\end{equation}
We put a  ``hat'' to distinguish this operator from the number of sites N.
Note that in all equations before we assumed summation over
momenta inside the magnetic  Brillouin zone. But in the eq. (\ref{totn}) we
must sum over the full Brillouin zone. Finally from eqs. (\ref{nn}),(\ref{totn})
one finds that the sum rule for the total number of electrons
\begin{equation}
\label{totsum}
\langle 0|\hat{N}|0\rangle = \sum_{\sigma, {\bf k}\in full}
{1\over 2} (1+2{{J}\over{t}}\gamma_{\bf k})=N
\end{equation}
is fulfilled. In conclusion of this discussion we would like to
note that the origin of all these complications with normalization
is very simple:The natural zone for the operator $d_{{\bf k}\sigma}$
is the magnetic Brillouin zone. On the other hand the natural zone
for $c_{{\bf k}\sigma}$ is the full Brillouin zone. This is the
reason why one should be careful comparing these two operators.

\section{Comparison with experiment}
Many of the experimental features observed are reproduced with the
theory described here. The large dispersion between $(0,0)$ and $(\pi/2,\pi/2)$
and the assymetric quasi particle weight about $(\pi/2,\pi/2)$ with the 
very strong dcrease in weight on moving beyond $(\pi/2,\pi/2)$. Also the
lack of dispersion along $(0,0)$ and $(\pi,0)$ as well as the very low 
quasiparticle weight is well reproduced. A major discrepancy between theory 
and experiment concerns the width of the quasi particle peak.The
theoretical spectra (Figs. 8,11) have narrow peaks. On the other hand
the widths of the experimental ARPES spectra\cite{Wells} are rather 
large, $\Gamma_0 \approx 0.3 eV \approx 2.4 J$.Although 
the experiment has been done at higher temperature $T=350K$ this is probably
too low to explain the widths of the peaks in terms of excited spin waves. 
It seems that another degree of freedom not considered in the present 
work must be of importance. One such possibility is the coupling to 
phonons. Such a coupling is expected to be quite large for the cuprates
especially for the Cu-O breathing mode since this strongly influences 
the stability of the Zang-Rice singlets. The coupling to phonons will
result in a broadening of the quasi particle peak which can be described 
in terms of the 
Franck - Condon factors describing the probability that the
system is left behind with a number of excited phonons
upon the sudden removal of one electron\cite{Saw,Alex}.
Such a large width would indicate a strong coupling with phonons.
In this paper we simulated the broadening with a
 Lorentz curve with a width $\Gamma_0$. The spectra obtained
by convolution of $-{1\over{\pi}} \ Im \ G_c^H(\epsilon, {\bf k})$
with the broadening curve are given in Fig. 12. Agreement of these spectra
with experimental ones\cite{Wells} is quite reasonable.

\section{Conclusions}
In the present work we consider a modified $t-J$ model at zero
doping (insulating copper oxide plane) and zero
temperature. We discuss the hole Green's functions with given pseudospin 
and given spin and derive a Dyson equation relating these two
Green's functions. The Green's function with given pseudospin is
very convenient for calculations, but it is an artificial object.
In the real experiment one measures the hole Green's function with given spin.

To describe the experiment we use hopping amplitudes for a modified $t-J$ model
as optained from LDA band calculations by Andersen {\it et al}\cite{And}.
Agreement of the theoretical spectra spectra with experimental 
ones\cite{Wells} is quite reasonable after some broadening of the theoretical
spectra.The physical origin of this broadening is not quite clear.
Possible explanations are in finite temperature,and the contributions due to 
phonons and the electron phonon interactions.

\section{Acknowledgments}
We are very grateful to D. Khomskii, D. C. Mattis and D. Poilblanc for
stimulating discussions, and A. Chernyshev for communicating
the results prior to publication.
One of us (O. P. S.) acknowledge the Laboratoire de Physique Quantique,
Universite Paul Sabatier; Materials Science Center, University of
Groningen; and Institute for Theoretical Atomic and
Molecular Physics at Harvard University (NSF grant PHY94-07194) for 
hospitality and support during work on the present problem.

\tighten

\newpage

Figure captions,\\

Fig.1 Plots of $-{1\over{\pi}} \ Im \ G_d(\epsilon, {\bf k})$
for different values of ${\bf k}$.
The Green's function is calculated in the self consistent Born approximation.\\

Fig.2  Quasiparticle residue $Z_{\bf k}^d$ of the hole Green's function
$G_d$. Solid line corresponds to the direction to the Brillouin zone
center: ${\bf k} \in \left[(\pi/2,\pi/2) \ - \ (0,0)\right]$.
Dashed line gives dependence along the face of the magnetic Brillouin zone: 
${\bf k} \in \left[(\pi/2,\pi/2) \ - \ (0,\pi)\right]$.\\

Fig.3 The vertices: a) - single hole creation, b) - hole + spin-wave
creation, c) - usual hole-spin-wave vertex.
Thick line correspond to $G_c$, and thin solid line corresponds to
$G_d$. Dashed line is spin-wave.\\

Fig.4 Hole production mechanisms. Upper part of the figure describes 
wave function of initial Neel state: a - component without spin quantum 
fluctuations, b - component with spin quantum fluctuation. Lower part of 
the figure arises instantly from the upper one after removal of an 
electron with spin up. 
Part a does not contain spin flip, and it corresponds to the amplitude 
$a_{\bf k}$. Part b does contain spin flip, and it corresponds to 
the amplitude $b_{\bf k,q}$.\\

Fig.5  Zero and first order in $t$ diagrams for hole photoproduction.\\

Fig.6  Dyson equation relating Green's functions $G_c$ (thick solid line)
and $G_d$ (thin solid line).\\

Fig.7 An example of higher order correction which is taken into account
in the Dyson equation (\ref{Dy2}).\\

Fig.8 Plots of $-{1\over{\pi}} \ Im \ G_c(\epsilon, {\bf k})$
for different values of ${\bf k}$.
The Green's function is found from Dyson equation (\ref{Dy2}) relating $G_c$
with that in self consistent Born approximation.\\

Fig.9 Dependence of quasiparticle residues on ${\bf k}$ 
along (1,1) direction.\\

Fig.10 Mechanisms of an electron removal without spin flip:
a) - $t-J$ model, b) - Hubbard model correction.\\

Fig.11 Plots of $-{1\over{\pi}} \ Im \ G_c^H(\epsilon, {\bf k})$
for different values of ${\bf k}$.
The Green's function is found from Dyson equation (\ref{Dy2}) relating $G_c^H$
with that in self consistent Born approximation.\\

Fig.12 Theoretical spectra for several values of ${\bf k}$. Spectra obtained 
by  convolution of  $-{1\over{\pi}} \ Im \ G_c^H(\epsilon, {\bf k})$ with 
spreading imitated by the Lorentz curve.

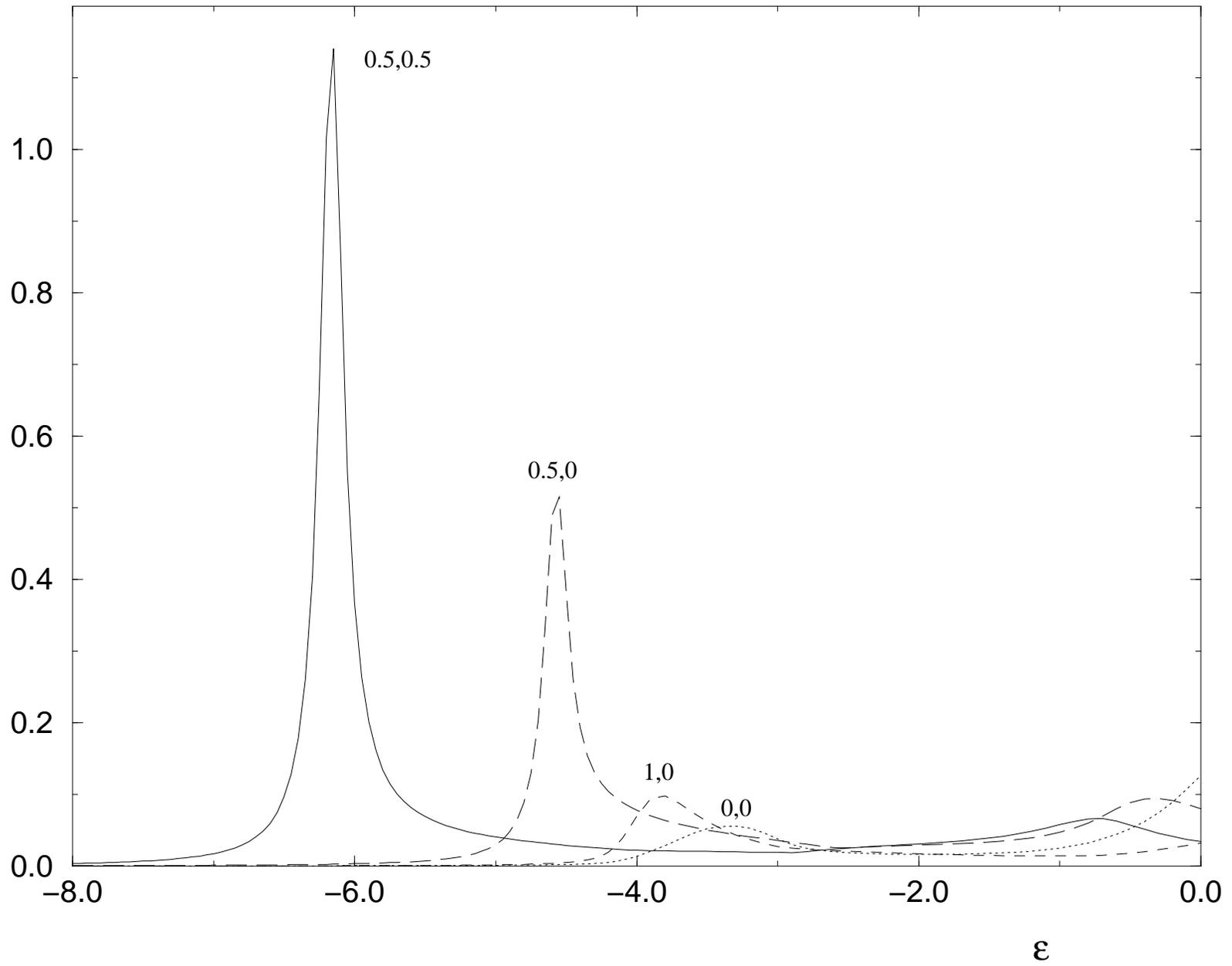
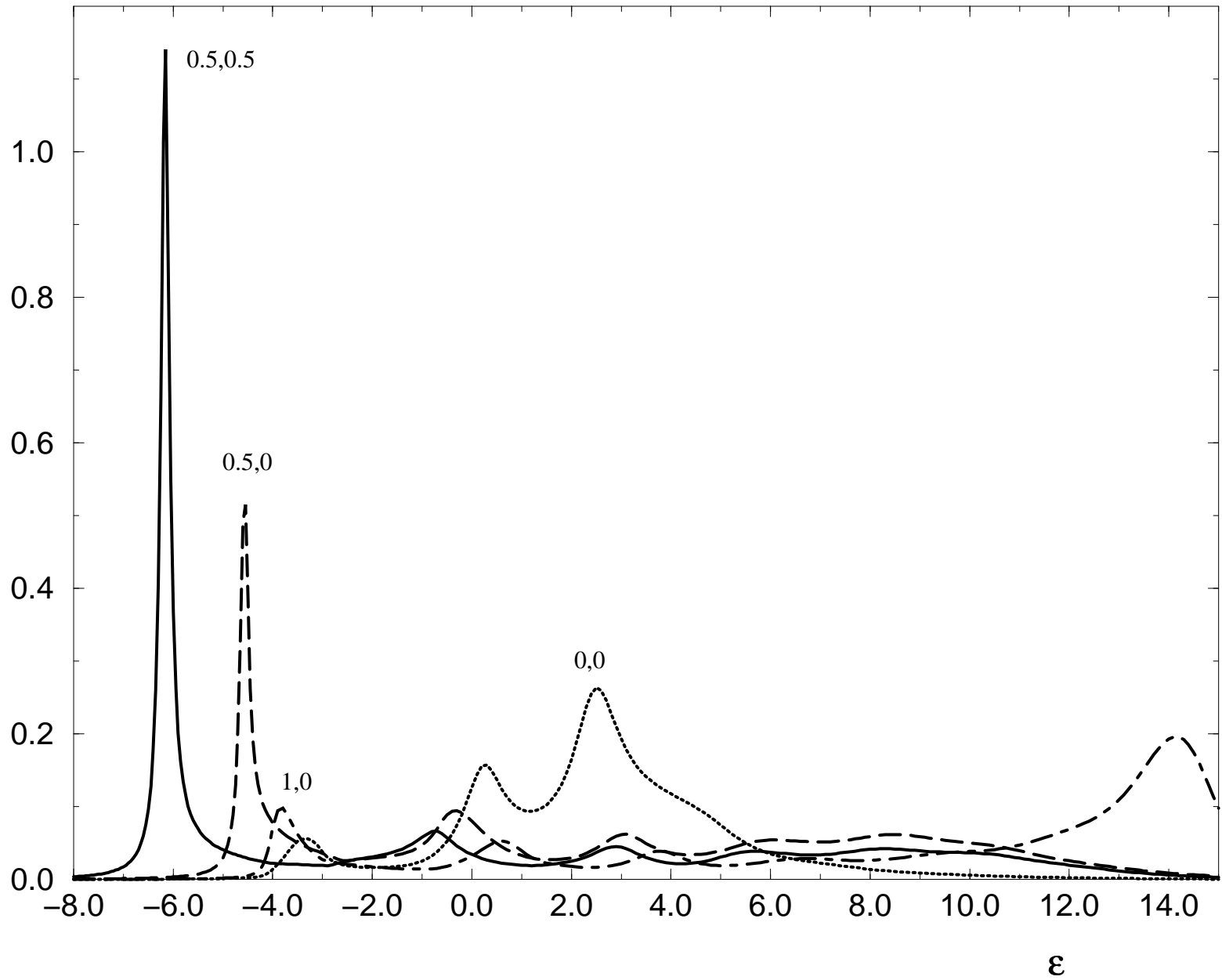
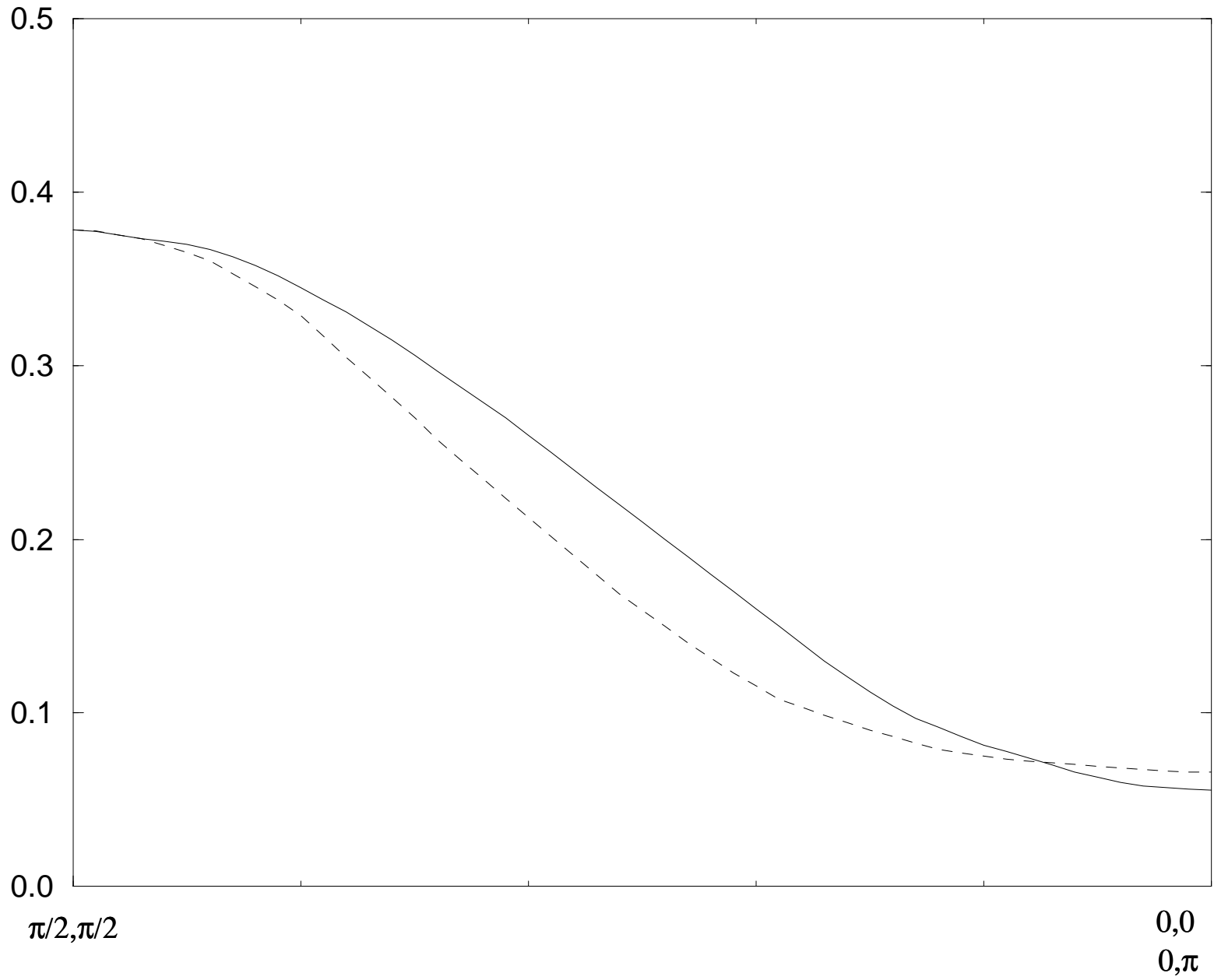
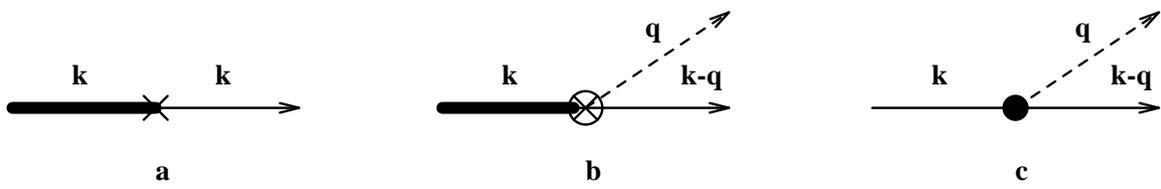
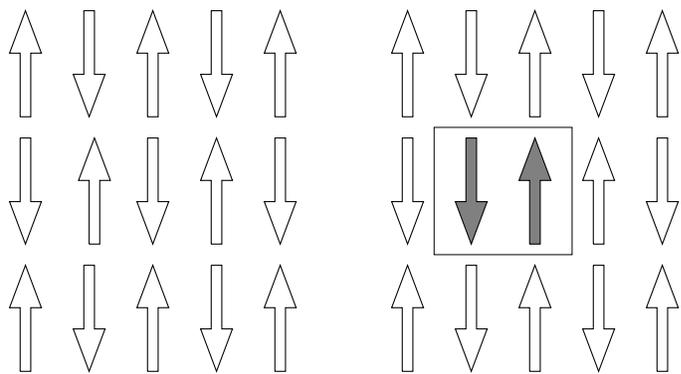
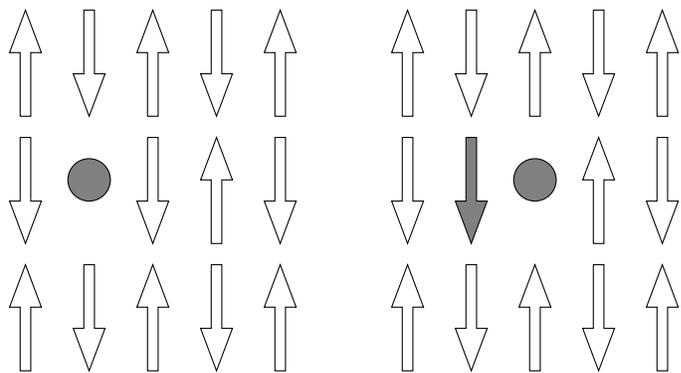
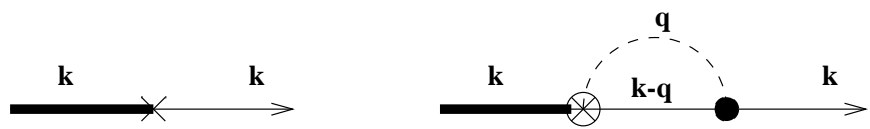
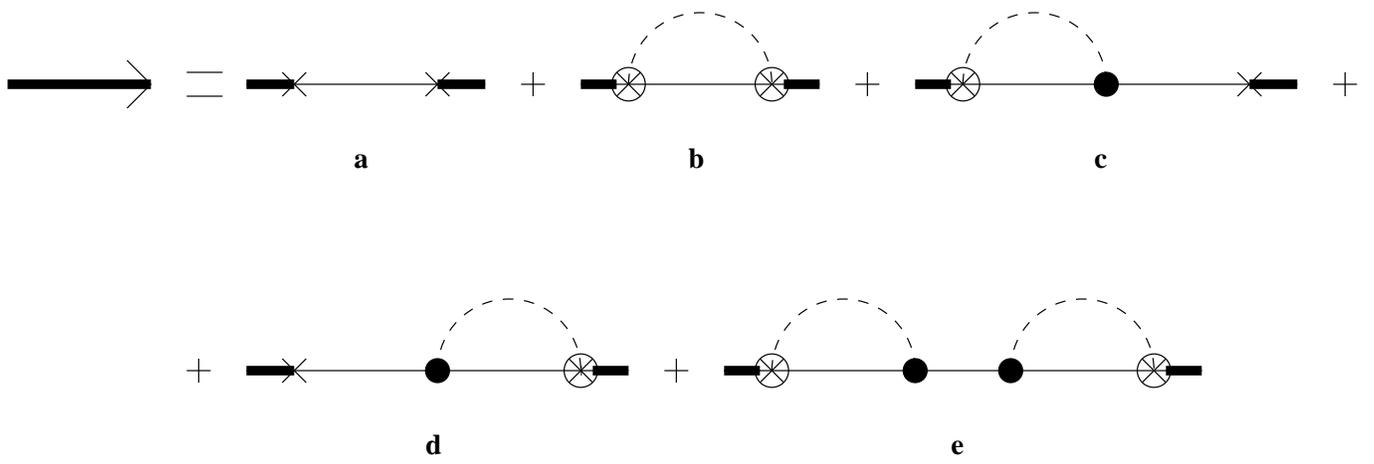
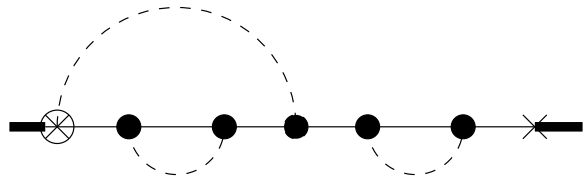
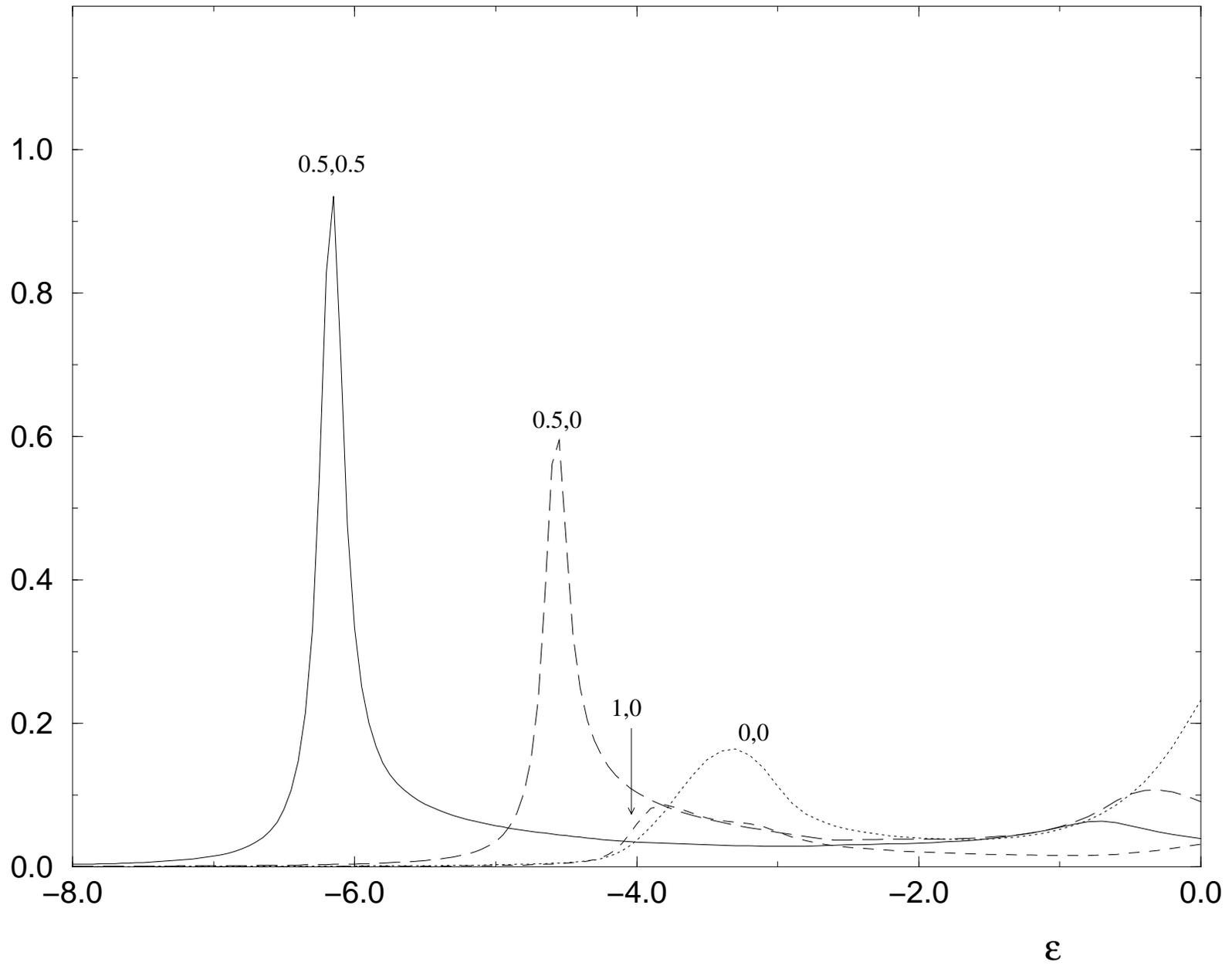
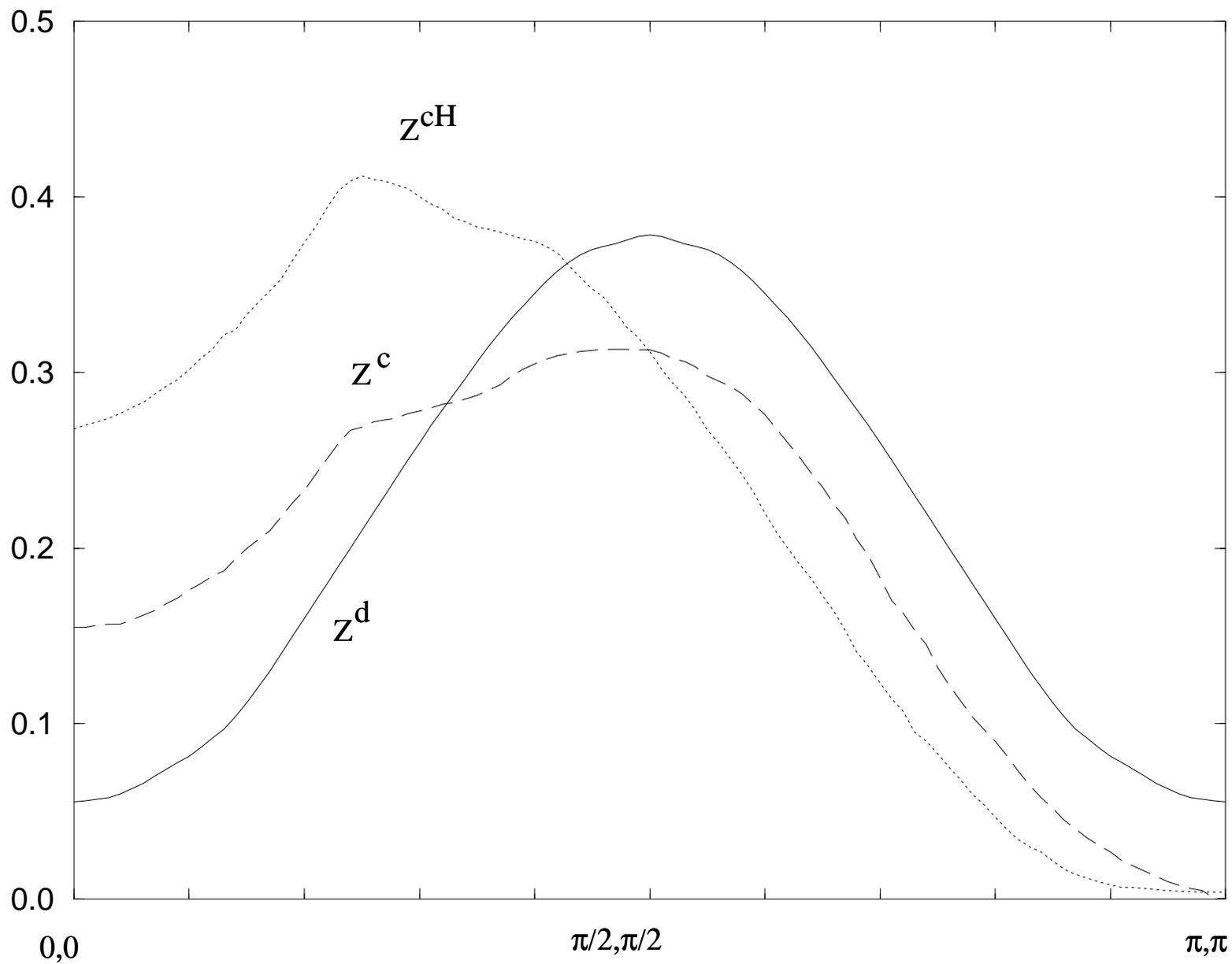
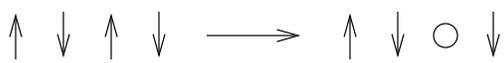
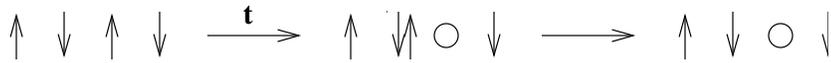
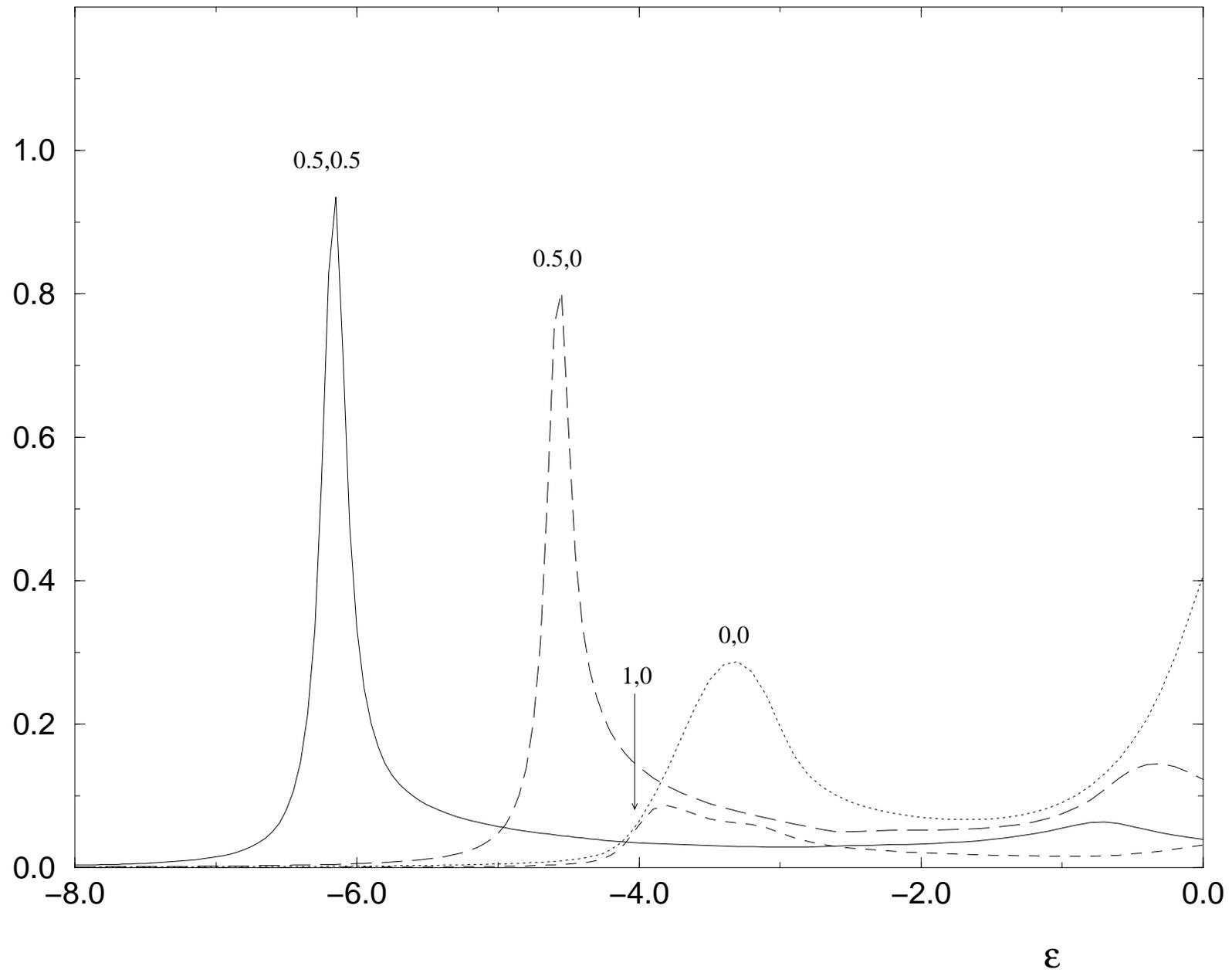
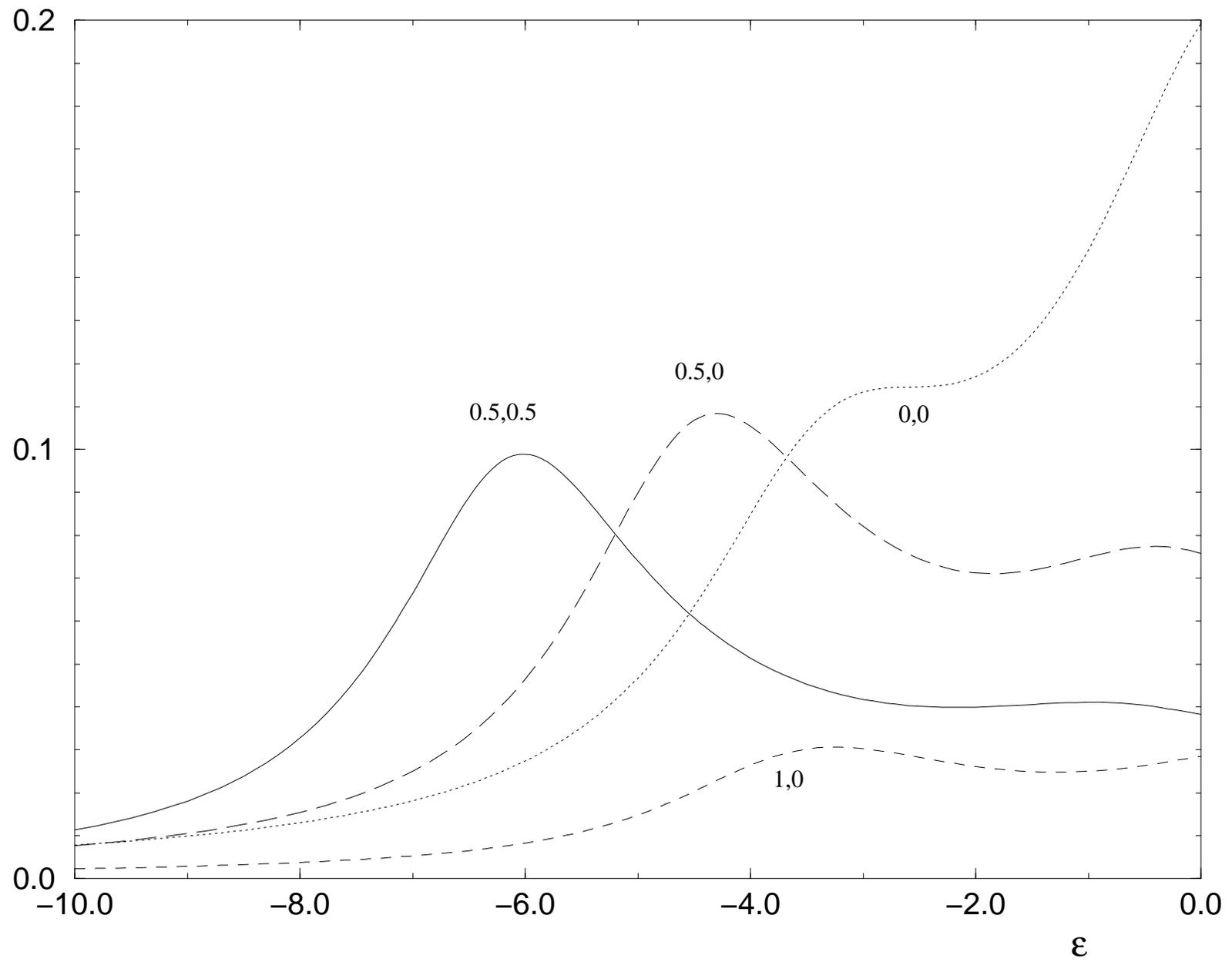
\end{document}